%% file: main.tex
\documentclass[11pt]{article}

\usepackage[final]{acl}

\usepackage{times}
\usepackage{latexsym}

\usepackage[T1]{fontenc}
\usepackage[utf8]{inputenc}

\usepackage{microtype}

\usepackage{inconsolata}

\usepackage{graphicx}

\usepackage{times}
\usepackage{latexsym}
\usepackage{amsmath}
\usepackage{amssymb}
\usepackage{float}
\usepackage{stfloats}
\usepackage{siunitx}
\usepackage{tabularx}
\usepackage{nicematrix}
\usepackage{makecell}
\usepackage{seqsplit}
\usepackage{booktabs}
\usepackage{caption}
\usepackage{longtable}
\usepackage{colortbl}
\usepackage{array}
\usepackage{anyfontsize}
\usepackage{multirow}
\usepackage{svg}
\usepackage{subcaption}
\usepackage[table]{xcolor}
\usepackage{url}
\usepackage{footmisc}
\usepackage{textcomp}
\usepackage{wasysym}
\usepackage{threeparttable}

\DeclareMathOperator{\tpk}{\textup{tp}_k}
\title{
  \LARGE Does generative AI supersede supervised XMLC? \\[0.5em]
  \Large A Benchmark Study on Automated Subject Indexing with German Scientific Literature
}

\author{Maximilian K\"ahler \and Katja Konermann \and Lisa Kluge
  \and Markus Schumacher\\
  Deutsche Nationalbibliothek \\
  Leipzig/ Frankfurt am Main, Germany\\
\\
 \small{
   \textbf{Correspondence:} \href{mailto:m.kaehler@dnb.de}{m.kaehler@dnb.de}
 }
}

\begin{document}
\maketitle
\begin{abstract}
  With a large controlled vocabulary as the label set, the task of automated
  subject indexing in a library can be understood as a multi-label 
  classification task. If the set of subject terms is large, the problem
  fits the Extreme Multi-Label Classification (XMLC) objective.
  In this study, we apply a selection of
  specialised supervised XMLC methods to the test case of subject indexing
  contemporary German scientific literature, collected at the German National Library (DNB).
  We contrast these results
  by including a classical lexical matching baseline and three of our own recently
  developed LLM-based methods into the benchmark.
  Algorithms are evaluated and compared in
  several metrics. This includes binary relevance comparisons with previously
  indexed material, as well as graded relevance ratings by professional subject
  librarians.
  A challenge for all methods is to reliably make suggestions from the long
  tail of the subject vocabulary. We find that supervised XMLC algorithms
  relying on transformer-based dense features give best results in terms of
  overall binary relevance metrics. However, focusing on graded relevance and
  performance in the long tail of our subject vocabulary, the LLM-based
  generative methods give better results, making them a promising alternative
  for future productive use.
\end{abstract}

\input{sections/01_introduction.tex}

\input{sections/02_datasets.tex}

\input{sections/03_methods.tex}

\input{sections/04_metrics.tex}

\input{sections/05_results.tex}

\input{sections/06-08_final-sections.tex}

\bibliography{benchmarkstudy}

\input{sections/App_datasets.tex}

\input{sections/App_examples.tex}

\input{sections/App_results.tex}

\input{sections/App_metrics.tex}

\end{document}

%% file: sections/01_introduction.tex
\section{Introduction}

Subject indexing is the task of assigning normed subject headings
from a controlled vocabulary to documents. In times of growing online repositories
and digital collections, its automation becomes a crucial task in modern library
processes \citep{Golub2021}.
At the German National Library (DNB), subject indexing is based on the
Integrated Authority File (GND). It is a shared authority file used in the
German-speaking countries. In its entirety, the GND holds 1.4 million potential
concepts, including not only subject headings but also individual person
names, geographic names, works, and other entity groups. Given its extreme
size, the GND—understood as a label set for multi-label classification—poses
a severe challenge. Not only is the label set large, but in addition, we
observe an extremely sparse and imbalanced label distribution.
Figure~\ref{fig:label_distribution} illustrates the GND's usage statistics
in the DNB's catalogue featuring a typical long-tail distribution:

\begin{figure}[h]
  \centering
  \includegraphics[width=0.48\textwidth]{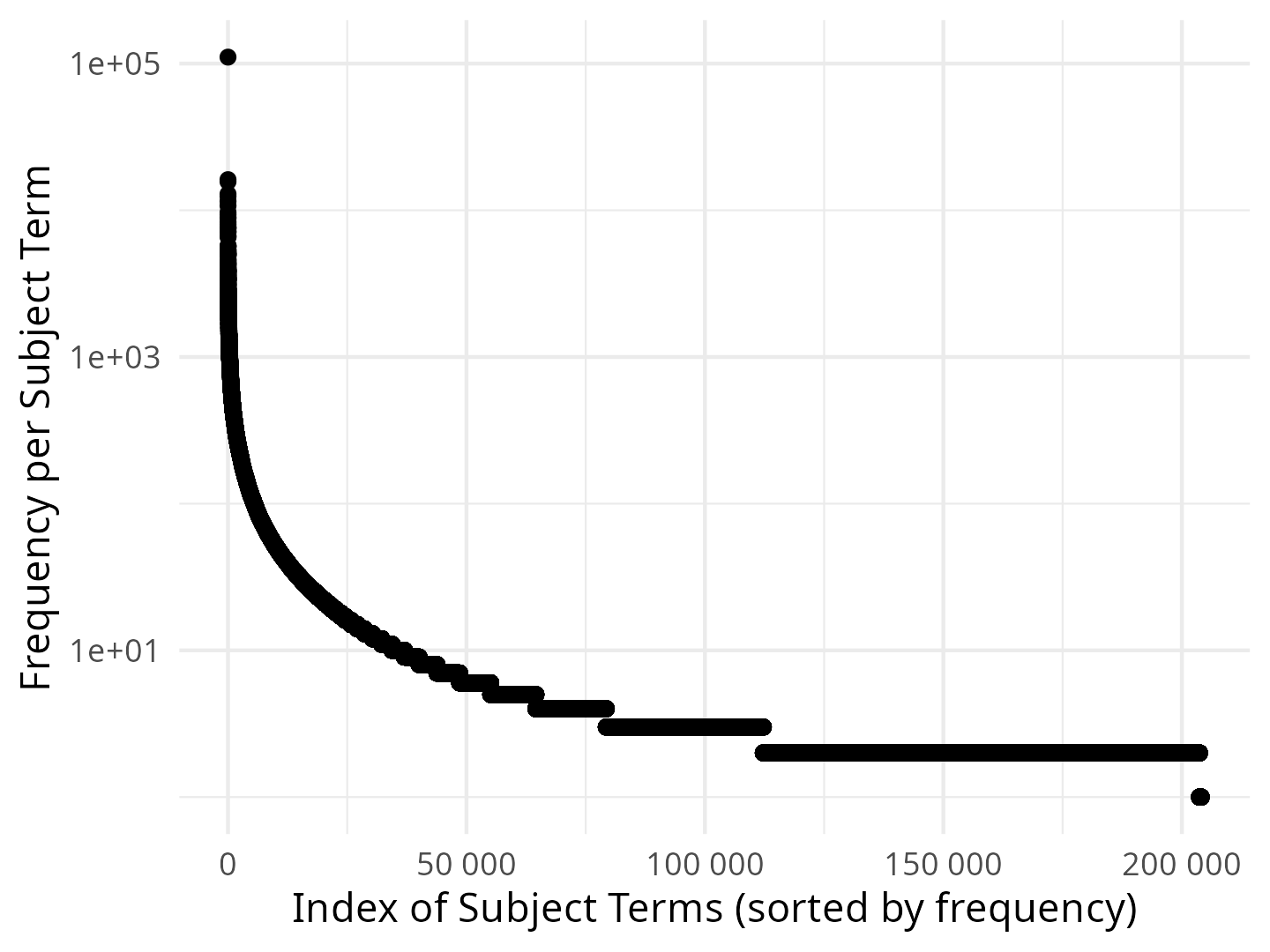}
  \caption{Distribution of subject term usage in the DNB catalogue (log scale).
  Only labels with at least one occurrence are shown.}
  \label{fig:label_distribution}
\end{figure}

These usage statistics are based on manual subject assignments conducted by
subject experts of the DNB. 
The process of manual subject assignment follows
the German rules for subject cataloguing
(RSWK\footnote{\url{https://www.dnb.de/rswk}}), and subject experts are trained
in these rules and possess in-depth subject knowledge in their respective
scientific domain.

Given the large, sparsely annotated, and highly imbalanced label set, the task
of automated subject indexing can be understood as an extreme multi-label
classification problem \citep{Bhatia16, PfastreXML}.
Two examples of manually indexed book titles can be found in table~\ref{tab:example-books} in the appendix.

\subsection{Related Work}

XMLC problems arise in a variety of applications: \citet{Parabel, Bonsai}
apply the paradigm to dynamic search advertising.
In \citet{X-Transformer, XR-Transformer}, XMLC methods are developed for
item-to-item recommendation, and \citet{Chalkidis2019a, Chalkidis2019b}
study the application to legal document classification.
Benchmark datasets for these applications are collected in the Extreme
Classification Repository \citep{Bhatia16}.
\citet{Wei2022, Dasgupta2023} put together a survey of XMLC methods.
Recently, \citet{DSouza2026} contributed a public benchmark dataset
containing multilingual library records also tagged with the GND.
In this benchmark, the top-performing team
\citep{suominen-etal-2025-annif, suominen-etal-2025-annif-germeval}
has applied a combination of XMLC methods and lexical matching,
using the Annif framework \citep{Suominen2019}.
The same kind of ensemble modelling is applied by productive services at the 
DNB \citep{poley-etal-2025}, too.

\subsection{Contributions of this Work}

This study is the first benchmark of XMLC algorithms on a purely German language
dataset. Furthermore, we apply algorithms not only to short texts but also to
longer texts, challenging their ability for long-context retrieval. In addition,
the evaluation is not only based on a binary relevance comparison with previously
indexed gold standard material, but also includes a graded relevance rating
conducted by subject experts on previously unseen material. Finally, the benchmark
of XMLC methods is complemented by the evaluation of traditional lexical matching
approaches as well as newly developed embedding-based matching techniques and
LLM-based few-shot prompting.

%% file: sections/02_datasets.tex
\section{Datasets}
\label{sec:data}

\subsection{Subject Vocabulary}

Our datasets restrict the subject indexing tasks to a subset of the Integrated
Authority File \texttt{GND-204K}, comprising 204\,056 distinct entities.
This collection includes
six entity types: Subject headings, geographic names, person names,
works, conferences, and corporate entities. Table~\ref{tab:gnd-stats} summarises the
numbers of distinct entities in each entity type. In the following, as a
wording of convenience, we define \textit{subject term} to mean any distinct
entity or concept from any of the entity groups, although this may not strictly
refer to the entity group of  \textit{subject headings} alone, but also includes
the other entity groups. For brevity, we also use the term \textit{label}
synonymously with the notion of subject term.
Furthermore, we refer to \texttt{GND-204K} as our \textit{label space} or
\textit{target vocabulary}.

 \begin{table}[!h]
  \small
  \centering
  \begin{tabular}{lr}
    \hline
      \textbf{GND Entity Type} & \textbf{Frequency} \\
    \hline
      Geographic Names &	24\,359 \\
      Corporate Entities &	22\,806 \\
      Person Names &	44\,998 \\
      Subject Headings &	94\,362 \\
      Conferences &	1\,249 \\
      Works &	16\,282 \\
    \hline
  \end{tabular}
    \caption{Entity types in the Integrated Authority File and their frequency of
  occurrence in the \texttt{GND-204K} subset of this benchmark study.}
    \label{tab:gnd-stats}
 \end{table}

The \texttt{GND-204K} subset is the collection of all entities that have been used
at least once in the available records of manually indexed contemporary
German scientific literature in our catalogue. The \texttt{GND-204K}
subset was created before splitting the available gold standard records into training,
development, and test sets (see the following sections).\footnote{Appendix~\ref{sec:add-stats}
provides further details on the splitting procedure.} Therefore, each of the
splits contains a portion of rare labels, not occurring in the other splits.
Labels occurring in the test set, but not in the training set, are referred to
as \textit{zero shot labels}.

\subsection{Training dataset}
\label{sec:trainingset}

We evaluate all algorithms on two tasks:

\begin{description}
  \item[Book-Titles] Subject indexing based on a document's mere title.
  \item[Fulltext-30k] Subject indexing based on the first 30\,000 characters of a
    document's full text.
\end{description}

All documents are textbooks of German scientific literature.

For each task there is a respective training and test set.
Table~\ref{tab:training-stats} shows some descriptive statistics of the respective
training datasets for each task. For development and hyper-parameter optimisation
we used additional splits \texttt{dev-train} and \texttt{dev-test} for each task.
Appendix~\ref{sec:add-stats} provides further details on our four-way-hold-out 
strategy. Details on how the test set was used for graded relevance evaluation
are explained in the following section~\ref{sec:evalset}.

\begin{table*}[!ht]
  \small
  \centering
  \begin{tabular}{lp{0.14\textwidth}p{0.14\textwidth}p{0.12\textwidth}p{0.12\textwidth}p{0.12\textwidth}}
    \hline
    \textbf{Task} & \textbf{Number of Documents} & \textbf{Distinct Labels} &
       $\diameter$\textbf{Labels per Record} &
       $\diameter$\textbf{Records per Label} &  
       \textbf{Bag-of-Words Feature Size}\footnotemark\\
    \hline
    \multicolumn{6}{l}{\textbf{Training set}} \\
    \hline
    \texttt{Book-Titles} & 951\,104 & 193\,921 & 2.76 & 13.5 & 650\,692\\
    \texttt{Fulltext-30k} & 167\,281 & 67\,757 & 3.37 & 8.32 & 1\,612\,482 \\
    \hline
    \multicolumn{6}{l}{\textbf{Test set}} \\
    \hline
    \texttt{Book-Titles} & 4\,651 & 8\,978  & 4.04 & 2.09 & 650\,692\\
    \texttt{Fulltext-30k} & 4\,651 & 8\,978 & 4.04 & 2.09 & 1\,612\,482 \\
    \hline
  \end{tabular}
  \caption{Statistics of the subject indexing datasets.}
  \label{tab:training-stats}
\end{table*}

Appendix~\ref{sec:length-distr} illustrates the distribution of text length for
the two tasks. In particular, figure~\ref{fig:length-distro-ft} illustrates the
length cut off at 30{,}000 characters. We note, that the cut off
at 30{,}000 characters constitutes a severe restriction in terms of contents used
in contrast to full length text books. 
However, ablation studies conducted with the Omikuji model 
(cf. section~\ref{sec:methods_omikuji}) in our productive 
services have indicated only a moderate loss of performance in terms of F1-Score. 
We hypothesise that the first 30{,}000 usually contain tables of contents or 
other summarizing parts of a book, that cover the books' wider content. 
Dealing with full length text books has been computationally out of scope for
this study.     

\footnotetext{While bag-of-words feature matrices are sparse in general, we observe
that the feature matrix for the \texttt{Book-Titles} task is considerably smaller in feature
size, as opposed to the \texttt{Fulltext-30k} task. }

\subsection{Evaluation dataset}
\label{sec:evalset}

Evaluation is conducted on a shared test set for both
tasks\footnote{Using a document's title for task \texttt{Book-Titles} or
  full text for task \texttt{Fulltext-30k} respectively},
consisting of 4\,651 documents equally distributed across 20 scientific subject
 groups. This ensures that all these subject categories
contribute with equal weight to the evaluation, even if the training data
is unevenly distributed over subject categories.\footnote{See appendix~\ref{sec:appendix}
for details on how we arrived at the various development sets to avoid optimisation bias.}
A graded relevance evaluation by subject experts was conducted on subsamples
of the test set.  Note, in particular, that the graded relevance evaluation is conducted
on distinct, non-overlapping, subsets of the test set so that each method was
rated on  500-600 documents, respectively. This is to ensure that each record is
 only seen once by every subject expert. Appendix~\ref{sec:appendix-sampling-plan} 
describes the details of our sampling plan for graded relevance evaluation.

% We compute boot-strap based confidence intervals for all metrics to illustrate
% uncertainty due to variation of the underlying data.

%% file: sections/03_methods.tex
\section{Methods}

Given the large variety of research prototypes for extreme multi-label
classification, this benchmark study selects only representatives of larger
families of algorithms. The selection was based on performance in the Extreme
Classification Repository Wiki500 benchmark \citep{Bhatia16} and popularity
in literature at the time, as measured by google scholar citations.
To have a baseline to our current library system, we included a lexical
indexing method. Finally, addressing the rising prevalence of embedding-based
methods and generative language models, we contrast the supervised XMLC methods
with three of our own recent prototypes.
Table~\ref{tab:method-comparison} summarises some important characteristics of
the algorithms included in this benchmark.

\begin{table*}[!ht]
  \centering
  \small
  \begin{tabular}{lcp{0.13\textwidth}p{0.15\textwidth}p{0.13\textwidth}}
    \hline
    \textbf{Method} & \textbf{Text Representation} & \textbf{Supervised Approach (y/n)} & \textbf{Zero-Shot Capabilities (y/n)} & \textbf{Usage of Label Features (y/n)} \\
    \hline
    \multicolumn{5}{l}{\textbf{Supervised XMLC Methods}} \\
    \hline
    DiSMEC++ & sparse (Bag-of-Words, TF-IDF) & \checkmark & \texttimes & \texttimes \\
    Omikuji & sparse (Bag-of-Words, TF-IDF) & \checkmark & \texttimes & \texttimes \\
    ZestXML & sparse (Bag-of-Words, TF-IDF) & \checkmark & \checkmark & \checkmark \\
    AttentionXML & dense (Transformer + LSTM) & \checkmark & \texttimes & \texttimes \\
    XR-Transformer & sparse \& dense (Transformer + TF-IDF) & \checkmark & \texttimes & \texttimes \\
    NGAME & dense (Transformer) & \checkmark & \texttimes & \checkmark \\
    \hline
    \multicolumn{5}{l}{\textbf{Unsupervised Methods}} \\
    \hline
    Lexical Matching & sparse (Bag-of-words) & \texttimes & \checkmark & \checkmark \\
    EBM & dense (Transformer) & \texttimes & \checkmark & \checkmark \\
     LLM-Ensemble & dense (Transformer) & \texttimes & \checkmark & \checkmark \\
    \hline
    KI-FSPrompt\footnotemark &
      dense (Transformer) & (\checkmark) & \checkmark & \checkmark \\
    \hline
  \end{tabular}
  \caption{Comparison of Methods Used in the Benchmark Study.}
  \label{tab:method-comparison}
\end{table*}
\footnotetext{While this is not strictly a supervised fine-tuning,
as, e.g., for the XR-Transformer approach, KI-FSPrompt draws strength from the
training material through the additional retrieval step.}

\subsection{Supervised XMLC Methods}

\subsubsection{DiSMEC++}

A common approach for multi-label tasks are 1vsAll classifiers, where a separate
classifier is trained for each label. Among others, \citet{DiSMEC++,DiSMEC}
propose one such algorithm, DiSMEC (\textit{Distributed Sparse Machines for
Extreme Multi-Label Classification}), and its successor DiSMEC++.
DiSMEC features a double layer of parallelisation for distributed computing,
and a capacity control mechanism to cut-off model weights to ensure compact model
size. In addition, DiSMEC++ further optimises the choice of initial weights,
resulting in an implicit negative mining effect. DiSMEC++ is designed to work
with a sparse TF-IDF feature matrix as representation of the input texts.

\subsubsection{Linear Models with Label-Partitioning
Techniques}
\label{sec:methods_omikuji}

While 1vsAll classifiers perform well,
training and run time scale linearly with the number of labels. As such,
naively employing them for problems with a large label set is often infeasible
for settings where efficient performance is necessary. To cut down on training
and inference time, \citet{Parabel} recursively cluster labels based
on their similarity. This creates a \textbf{partitioned label tree} (PLT),
in which inner nodes can be considered metalabels, and node classifiers decide whether to
follow a child node. As negative examples for training a classifier can be
efficiently selected from a node, efficient training is possible. During
inference, beam search can be employed to avoid evaluating all node
classifiers.\\
While \citet{Parabel} use binary trees with balanced label clusters,
\citet{Bonsai} propose Bonsai, where a node can have more than
two children, which leads to more shallow PLT. Additionally, Bonsai does not
enforce the same size between all clusters.\\
For our experiments, we use
\texttt{Omikuji}\footnote{\href{https://github.com/tomtung/omikuji}{https://github.com/tomtung/omikuji}},
which implements various PLT methods with linear classifiers,
re-implementing works of \citet{Parabel},
\citet{Bonsai} and \citet{AttentionXML} in the RUST language. Similar to DiSMEC++, Omikuji operates on
sparse TF-IDF feature matrices to process input documents.

\subsubsection{ZestXML}

Another approach working on sparse TF-IDF feature matrices is ZestXML
(\textit{Generalized Zero-Shot Extreme Multi-label Learning}) by \citet{ZestXML}.
In addition to learning correlations between TF-IDF-text features and the
document-label matrix, as done by all of the above-mentioned XMLC methods,
this algorithm also represents labels through their TF-IDF
features and correlates text features with label features directly. Another
interesting feature is the ability to incorporate a label hierarchy in the
label features and add this to the learnable features. As such, ZestXML has
the ability to also infer on Zero-Shot labels, if some of their label features
have been seen in the training material.

\subsubsection{AttentionXML}

Earlier approaches for XMLC usually rely on bag-of-words representations of
input documents, like TF-IDF matrices. AttentionXML by
\citet{AttentionXML} employs an LSTM, a recurrent deep-learning
architecture, for XMLC, which is able to capture more context-dependent relations
within an input document. Additionally, a label-wise attention mechanism is
employed, which allows for separate input representation for each label.
\citet{AttentionXML} also cluster labels in a PLT for efficient
training and inference, but in contrast to approaches like
\citet{Parabel} and \citet{Bonsai}, AttentionXML
trains classifiers in a layer-wise fashion, where neural networks in deeper
layers of the PLT are initialised with parameters of the previous layer for
fast convergence. Like \citet{Bonsai}, the authors opt for
a more shallow label tree, which is achieved via layer collapsing.\\
In our experiments, we found initial performance to be low due to a high number
of unknown tokens when using the originally proposed \texttt{fasttext} word
tokenisation. Instead, we use Byte-Pair Encoding tokenisation, which splits
unknown words into previously encountered subwords. We then initialise
(sub-)word embeddings from a BERT embedding layer.

\subsubsection{XR-Transformer}

The Transformer architecture with a self-attention mechanism has been long
established as state-of-the-art for many natural language processing tasks
\citep{Vaswani2017}. Transformer models like BERT
\citep{devlin_bert_2019} are usually pre-trained in an unsupervised manner on
large text corpora before they are fine-tuned on task-specific datasets.
\citet{XR-Transformer, X-Transformer} adapt transformers for extreme multi-label
classification. Similar to previously discussed approaches, a label tree is
employed. A pre-trained encoder model is fine-tuned layer-wise on the meta-label
classification tasks. For the final predictions, linear classifiers are trained
on the concatenation of the fine-tuned, dense Transformer embeddings and the
sparse TF-IDF vectors.

\subsubsection{NGAME}

Many approaches for XMLC, like by \citet{AttentionXML} and
\citet{XR-Transformer}, treat labels as mere identifiers. However, labels are
often associated with textual data. As such, this side-information can be
leveraged for extreme multi-label classification, as also attempted by
\citet{ZestXML}. NGAME \cite{NGAME}
makes use of textual data associated with labels by embedding both input
documents and label texts with a pre-trained encoder model, and minimizing the
distance between the embeddings of input documents and their associated labels.
A challenge for this set-up is the selection of informative hard negative
examples to train on: \citet{NGAME} employ curriculum learning,
where batches are constructed based on clusters with similar embeddings. Negative
examples can be efficiently sampled from the same batch. As the cluster size
decreases during training, examples in a batch become more similar, and as such
more difficult negative labels are introduced.

\subsection{Unsupervised Approaches}

\subsubsection{Lexical Matching}

Subject indexing can also be approached as an instance of keyphrase 
extraction~\citep{erbs2013bringing}. A special case is keyphrase extraction
using lexical matching \citep{Frank99domain-specifickeyphrase}. In
\citet{Medelyan2008, Maui}, the algorithm Maui was introduced. Maui combines
the lexical matching with a candidate ranking, based on heuristics acquired
during the lexical matching process: How often does the match occur in a text?
What was its position in the text? What is its distribution throughout the document?
These heuristics are used to train a small ranker model with a much smaller
training dataset than in the above-mentioned supervised XMLC methods.\footnote{
  We classify this algorithm as unsupervised, as it is a mere fraction of the
  training data needed to train the ranker in such approaches.}
The Annif toolkit also implements the idea of lexical matching in its backend
MLLM (\textit{Maui Like Lexical Matching}) \citep{Suominen2021}, which we
include as a baseline in our benchmark.

\subsubsection{Embedding Based Matching (EBM)}

The idea of lexical matching can be ``modernised'' by relying on embedding-based
matching instead of lexical matching. This is closely related to the idea of dense retrieval models \citep{DRMSurvey}. Both input documents and vocabulary
can be vectorised by using a pre-trained sentence encoder model, e.g.,
\texttt{BGE-m3}~\citep{bge-m3} or \texttt{jina-embeddings-v3}~\citep{JinaAI}.
Vector search across the vocabulary creates candidate matches. As in Maui or
MLLM, these candidate matches can be reranked using a small ranking model
trained on similar heuristics that are captured during the matching process
(frequency of occurrence, position, spread). This idea is implemented in the
\texttt{ebm4subjects} package~\citep{Khler2025a}.

\subsection{Methods using generative LLMs}

\subsubsection{LLM-Ensemble}

The LLM-Ensemble \citep{Kluge2025} approaches the subject indexing problem by
combining multiple LLMs with multiple fixed few-shot prompts. The LLM-generated
subject terms are mapped to the vocabulary via hybrid search, and further
post-processing steps combine the individual suggestions to an ensemble vote.
This approach only needs a small number of examples filling the few-shot
prompts, so there is no need for training. However, calling multiple
LLMs with multiple prompts per document comes with high inference costs. In
particular, longer texts that are processed chunkwise require many LLM-calls
at inference time. Language models included in the LLM-Ensemble for completing few-shot
prompts are \texttt{Llama\--3.2\--3B\--Instruct}, \texttt{Llama\--3.1\--70B\--Instruct},
\texttt{Mistral\--7B\--v0.1}, \texttt{Mistral\--7B\--Instruct\--v0.3},
\texttt{Mixtral\--8x7B\-Instruct\--v0.1}, \texttt{OpenHermes\--2.5\--Mistral\--7B}, and
\texttt{Teuken\--7B\--instruct\--research\--v0.4}. In addition,
\texttt{Llama\--3.1\--8B\--Instruct} is used as ranking model.

\subsubsection{KI-FSPrompt}

KI-FSPrompt \citep{Khler2025} is another system that leverages the idea of
few-shot prompting, to generate subject terms with a generative LLM.
As an attempt to mitigate the high inference costs of the LLM-Ensemble, the
system KI-FSPrompt alters the approach of the LLM-Ensemble
by replacing the fixed few-shot prompts with a per-document RAG-like retrieval
step, where few-shot prompts are assembled dynamically with examples from the
training dataset that are close to the requested document (in terms of
embedding similarity). We use
KI-FSPrompt with \texttt{Mistral-7B-Instruct-v0.3} and \texttt{Llama-3.2-3B-Instruct}
as models for prompt-completion, and \texttt{Meta-Llama-3.1-8B-Instruct} as
a ranking model.

%% file: sections/04_metrics.tex
\section{Metrics}
\label{sec:metrics}

In this benchmark study, we use binary relevance and graded relevance metrics.

\subsection{Binary Relevance}

For binary relevance, we compare the machine-based subject suggestions with
the gold standard, annotated by subject experts according to the German rules
for subject cataloguing (RSWK).
All algorithms studied in this benchmark produce a list of subject
suggestions, where every subject suggestion comes with some confidence score
that can be used to obtain a ranked list of subject suggestions per document.
Therefore, standard metrics for ranked retrieval apply.
However, in practice, one would usually filter predictions with some uniform limit $k$ on
ranks and threshold $t$ on confidence scores to filter out the most relevant
suggestions for a productive system. Focus on high recall or high precision
depends on the use case: an unsupervised cataloguing system needs to be
calibrated more strictly to achieve high precision; in a supervised (assisted)
workflow, producing high recall is more important.
To measure the performance over the entire ranking, we look at precision-recall
curves: for any given limit $k$ and threshold $t$, we compute the document average
(doc-avg) precision and recall. For every level of recall, there is an
optimal configuration
of limit and threshold achieving highest precision. This defines the
precision-recall
curve ($\mathcal{C}_\textup{pr}$). Computing the integral of the pr curve gives rise to the
measure of the
\textbf{Area under the Precision-Recall Curve} ($\textup{AUC}_{\textup{pr}}$).
$\textup{AUC}_{\textup{pr}}$ will be our primary measure for algorithmic performance.
To give some point estimates of a system's performance, we also compare their optimal
(document average) $F^1$-score, denoted as $F^{1,*}$, that could be attained by applying
an optimal calibration of limit and threshold to the system. See appendix~\ref{sec:mediating-in-sample-bias}
for details on how we avoid in-sample bias in this estimation and see
appendix~\ref{sec:app-metrics} for full definitions of all metrics. These can
be computed using our public R package \texttt{CASIMiR}\footnote{\url{https://github.com/deutsche-nationalbibliothek/casimir}}.

\subsection{Binary Relevance with Conditional Weighting}

In the evaluation of subject indexing methods, it is very important to also take
performance on long-tail labels into consideration. Otherwise,
algorithms that simply ignore rare labels and focus on head labels will dominate
rankings. However, in large controlled subject vocabularies, the nuanced, most specialised topics
are those of highest interest for users in search of specific literature.
To study the performance in the long-tail of our vocabulary, we introduce
weighted variants of the above metrics that use a specific cost scheme.
True positives ($\text{tp}$) and false negatives ($\text{fn}$)
are weighted by a label-dependent cost factor that is based on inverse
propensity scores \citep{PfastreXML}. We denote these
\textbf{conditionally propensity scored metrics} as
$\textup{cps-Rec}, \textup{cps-Prec}, \textup{cps-F}^1$ and $\textup{cps-AUC}_\textup{pr}$,
 and report them separately. See appendix~\ref{sec:app-metrics}
for details.

\subsubsection{Graded Relevance}

Whereas binary relevance comparison treats all false positive suggestions as
simply wrong, the graded relevance rating allows comparing to what degree
false positive suggestions still constitute helpful subject assignments, even if
they would not be assigned after RSWK rules.
In the graded relevance
scenario, subject experts directly rate the top 5 machine-based subject suggestions
for each algorithm and task. They use
an ordinal scale of (0) false, (1) slightly useful, (2) useful and (3)
very useful. Here, usefulness is defined by a subject term's suitability as
a search query to find a particular document in the catalogue. This is a
notion of usefulness that takes the user's perspective into account, as opposed to strict adherence
to the German rules for subject cataloguing, which is measured by the binary relevance
comparison with the gold standard. To compare the algorithms in this graded relevance
scenario, we use the metrics \textbf{generalised precision} $\textup{g-Prec}$ and
\textbf{generalised recall} $\textup{g-Rec}$ as defined in \citet{Kekalainen2002}.
See appendix~\ref{sec:app-graded-relevance} for further details.

%% file: sections/05_results.tex
\section{Results}
\label{sec:results}

\subsection{Binary Relevance Results}
\label{sec:results-bin-rel}

\input{tables/table_bin_rel.tex}

Table~\ref{tab:bin_rel} shows the results of the binary relevance comparison
for both tasks. The ranking of methods in terms of $\textup{AUC}_{\textup{pr}}$
and optimal $F_1$-score $F^{1,*}$ is not consistent between the two metrics.
Notably, the supervised XMLC methods,
XR-Transformer, Omikuji, and DiSMEC++, lead the ranking for both tasks in terms
of $\textup{AUC}_{\textup{pr}}$.
Interestingly, while XR-Transformer outperforms the other methods by a significant
margin for the \texttt{Book-Titles} task, the difference between DiSMEC++ and
XR-Transformer is hardly notable for \texttt{Fulltext-30k}.
We hypothesise that, due to the relatively short context length of the finetuned
transformer, little is to be gained from the transformer part here, and most
predictive power is coming from the TF-IDF features alone.
The AttentionXML algorithm, not relying on bag-of-words features at all, shows
complete failure in the \texttt{Fulltext-30k} task: only little information can
be gained from looking at the very small context window allowed by
AttentionXML when looking at a full text.\footnote{The first 2–3 sentences of a
textbook are usually of formal nature and cannot contribute significantly to the
comprehension of the book's content.} The title alone
is more informative for AttentionXML's LSTM architecture.
While most methods achieve higher results in the \texttt{Book-Titles} task, we
see the opposite for the ontology-based matching methods, MLLM and EBM. Both
methods benefit from processing more text material.
Neither in the \texttt{Book-Titles} task nor in the \texttt{Fulltext-30k} task
do the supervised methods that make use of the label features (NGAME, ZestXML)
show distinguished results.

Looking at the conditionally propensity scored metrics in table~\ref{tab:bin_rel},
emphasising the long-tail performance of the methods, we see a slightly
different ranking.
The LLM-based KI-FSPrompt is ahead of XR-Transformer for the \texttt{Book-Titles}
task in the point estimate of $\textup{cps-}F^{1,*}$, but not
$\textup{cps-AUC}_{\textup{pr}}$.
The precision-recall curve in figure~\ref{fig:pr_curve_1} in the
appendix~\ref{sec:app-results} illustrates the situation.
In the \texttt{Fulltext-30k} task, the LLM-Ensemble
is clearly on top of the other methods, followed by XR-Transformer and DiSMEC++.
In these weighted metrics, even the lexical matching and embedding-based
matching can best the supervised partitioned label tree approach, Omikuji.
We illustrate these findings in appendix~\ref{sec:app-label-freq},
showing estimates of the subject average $F^1$-score by label frequency.

\subsection{Graded Relevance Results}
\label{sec:results-grad-rel}

\input{tables/table_graded_rel.tex}

Table~\ref{tab:graded_rel} shows the results of the graded relevance ratings
in terms of generalised precision, recall, and $F_1$-score at limit $k=5$.
We observe that the LLM-Ensemble method is ahead of all other methods in both
tasks.
Figure~\ref{fig:qm-ratings} in appendix~\ref{sec:app-results} offers some
explanation: While the portion of subject-term predictions rated as "very useful"
is only marginally larger than those generated by other methods, there is a
much larger portion of "useful" and "slightly useful", so that the
LLM-Ensemble does not produce as many "wrong" predictions as the other methods.
Another takeaway is that in the \texttt{Book-Titles} task, the supervised XMLC
methods with dense feature representations (Transformer/LSTM), namely
XR-Transformer and AttentionXML, score higher than the TF-IDF-based Omikuji and
ZestXML. This is not the case in the \texttt{Fulltext-30k} task.
In terms of generalised recall, the EBM method almost reaches the performance
of supervised methods for the \texttt{Fulltext-30k} task, which would qualify
EBM as a suitable method for assisted subject indexing workflows with little
training material.

\subsection{Inference Time}
\label{sec:results-inference-time}

Due to the complexity of the benchmarking process and changing hardware setups
during the course of the project, this project did not 
maintain comparable records of training and inference times for all algorithms in 
both tasks. Nevertheless, this is a vital component of evaluation. We therefore 
include a comparison of inference times for the \texttt{Book-Titles} task
computed on a 2 x Intel(R) Xeon(R) Gold 6338T CPU @ 2.10GHz (48 core) architecture
with two NVIDIA A100 80GB GPU processors. Table~\ref{tab:inference-time} displays
the inference times. 
% \footnote{Unfortunately, for some methods comparable 
%measurements are missing.} 

\begin{table}[!htb]
  \small
  \begin{threeparttable}
    \begin{tabular}{lr}
      \hline
      \textbf{Method} & \textbf{Inference Time p. rec. in ms} \\
      \hline
      DiSMEC++\tnote{*} & <1 \\
      Omikuji\tnote{*} & <1 \\
      AttentionXML\tnote{\textdagger} & 3 \\
      XR-Transformer\tnote{\textdagger} & 10.5 \\
      NGAME\tnote{\textdagger} & 1.6 \\
      ZestXML\tnote{*} & <1 \\
      KI-FSPrompt\tnote{\textdagger} & 271 \\
      LLM-Ensemble\tnote{\textdaggerdbl} & 998 \\
      Emb.-based matching\tnote{\textdagger} & 9.4 \\
      Lexical Matching (MLLM) & 23 \\
      \hline
    \end{tabular}

    \begin{tablenotes}[flushleft]
      \footnotesize
      \item[*] 40 parallel CPU-processes
      \item[\textdagger] 1 NVIDIA A100 GPU
      \item[\textdaggerdbl] Setup requires 2 NVIDIA A100 GPU 
    \end{tablenotes}
  \end{threeparttable}
  \caption{Inference times (wall time) for task \texttt{Book-Titles}}
  \label{tab:inference-time}
\end{table}

We observe that LLM-based methods are more resource-intensive by at least two
orders of magnitude, compared to the other methods. In particular, the
LLM-Ensemble, requiring several calls per record to larger 56B to 70B models,
does not only raise the demand for appropriate hardware (two GPUs required),
but reaches processing times that must be judged as infeasible for processing
large amounts of data in daily library processes. 
While for practical reasons we do not include a similar table for the 
\texttt{Fulltext-30k}, the inference times in this case are, of course, much higher
for all methods. Particularly,  LLM-Ensemble, Emb.-based matching and 
Lexical Matching (MLLM) scale linearly with text length. For Emb.-based matching
the generation of embeddings for longer texts can become a severe bottleneck.

%% file: tables/table_bin_rel.tex
\begin{table*}[t]
\small
\begin{tabular*}{\linewidth}{@{\extracolsep{\fill}}crrrrrrrr}
\toprule
 & \multicolumn{4}{c}{{Book-Titles}} & \multicolumn{4}{c}{{Fulltext-30k}} \\ 
\cmidrule(lr){2-5} \cmidrule(lr){6-9}
Method & $F^{1,*}$ & $\textup{AUC}_\textup{pr}$ & $\textup{cps-}F^{1,*}$ & $\textup{cps-AUC}_\textup{pr}$ & $F^{1,*}$ & $\textup{AUC}_\textup{pr}$ & $\textup{cps-}F^{1,*}$ & $\textup{cps-AUC}_\textup{pr}$ \\ 
\midrule\addlinespace[2.5pt]
DiSMEC++ & 0.426 & 0.402 & 0.386 & 0.368 & 0.361 & 0.354 & 0.305 & 0.306 \\ 
Omikuji & 0.417 & 0.395 & 0.370 & 0.354 & 0.343 & 0.329 & 0.277 & 0.273 \\ 
AttentionXML & 0.367 & 0.342 & 0.298 & 0.282 & 0.188 & 0.124 & 0.124 & 0.083 \\ 
XR-Transformer & {\bfseries 0.467} & {\bfseries 0.427} & 0.421 & {\bfseries 0.389} & {\bfseries 0.398} & {\bfseries 0.359} & 0.339 & 0.307 \\ 
NGAME & 0.391 & 0.366 & 0.383 & 0.369 & NA & NA & NA & NA \\ 
ZestXML & 0.379 & 0.369 & 0.359 & 0.356 & 0.336 & 0.295 & 0.301 & 0.267 \\ 
KI-FSPrompt & 0.453 & 0.340 & {\bfseries 0.436} & 0.339 & NA & NA & NA & NA \\ 
LLM-Ensemble & 0.382 & 0.308 & 0.379 & 0.316 & 0.364 & 0.333 & {\bfseries 0.362} & {\bfseries 0.345} \\ 
Emb.-based matching & 0.285 & 0.212 & 0.274 & 0.214 & 0.309 & 0.279 & 0.295 & 0.274 \\ 
Lexical Matching (MLLM) & 0.274 & 0.148 & 0.266 & 0.146 & 0.303 & 0.268 & 0.277 & 0.253 \\ 
\bottomrule
\end{tabular*}
\caption{Binary relevance results in Tasks \texttt{Fulltext-30k} and \texttt{Book-Titles}.}
\label{tab:bin_rel}
\end{table*}

%% file: tables/table_graded_rel.tex
\begin{table*}[t]
\small
\begin{tabular*}{\linewidth}{@{\extracolsep{\fill}}>{\raggedright\arraybackslash}m{\dimexpr 0.14\linewidth -2\tabcolsep-1.5\arrayrulewidth}m{\dimexpr 0.06\linewidth -2\tabcolsep-1.5\arrayrulewidth}>{\raggedright\arraybackslash}m{\dimexpr 0.07\linewidth -2\tabcolsep-1.5\arrayrulewidth}>{\raggedleft\arraybackslash}m{\dimexpr 0.06\linewidth -2\tabcolsep-1.5\arrayrulewidth}>{\raggedright\arraybackslash}m{\dimexpr 0.07\linewidth -2\tabcolsep-1.5\arrayrulewidth}>{\raggedleft\arraybackslash}m{\dimexpr 0.06\linewidth -2\tabcolsep-1.5\arrayrulewidth}>{\raggedright\arraybackslash}m{\dimexpr 0.07\linewidth -2\tabcolsep-1.5\arrayrulewidth}>{\raggedleft\arraybackslash}m{\dimexpr 0.06\linewidth -2\tabcolsep-1.5\arrayrulewidth}>{\raggedright\arraybackslash}m{\dimexpr 0.07\linewidth -2\tabcolsep-1.5\arrayrulewidth}>{\raggedleft\arraybackslash}m{\dimexpr 0.06\linewidth -2\tabcolsep-1.5\arrayrulewidth}>{\raggedright\arraybackslash}m{\dimexpr 0.07\linewidth -2\tabcolsep-1.5\arrayrulewidth}>{\raggedleft\arraybackslash}m{\dimexpr 0.06\linewidth -2\tabcolsep-1.5\arrayrulewidth}>{\raggedright\arraybackslash}m{\dimexpr 0.07\linewidth -2\tabcolsep-1.5\arrayrulewidth}>{\raggedleft\arraybackslash}m{\dimexpr 0.07\linewidth -2\tabcolsep-1.5\arrayrulewidth}}
\toprule
 & \multicolumn{6}{>{\centering\arraybackslash}m{\dimexpr 0.39\linewidth -2\tabcolsep-1.5\arrayrulewidth}}{\parbox{\linewidth}{\centering {Book-Titles}}} & \multicolumn{6}{>{\centering\arraybackslash}m{\dimexpr 0.39\linewidth -2\tabcolsep-1.5\arrayrulewidth}}{\parbox{\linewidth}{\centering {Fulltext}}} &  \\ 
\cmidrule(lr){2-7} \cmidrule(lr){8-13}
Method & $\textup{g-}F^{1}_5$ & {95\% CI} & $\textup{g-Prec}_5$ & {95\% CI} & $\textup{g-Rec}_5$ & {95\% CI} & $\textup{g-}F^{1}_5$ & {95\% CI} & $\textup{g-Prec}_5$ & {95\% CI} & $\textup{g-Rec}_5$ & {95\% CI} & $N_{\textup{docs}}$ \\ 
\midrule\addlinespace[2.5pt]
AttentionXML & \centering 0.523 & {[0.514, 0.536]} & 0.533 & {[0.526, 0.547]} & 0.565 & {[0.553, 0.584]} & 0.337 & {[0.331, 0.353]} & 0.344 & {[0.337, 0.361]} & 0.361 & {[0.354, 0.375]} & 524 \\ 
Emb.-based matching & 0.433 & {[0.420, 0.443]} & 0.440 & {[0.425, 0.449]} & 0.473 & {[0.463, 0.487]} & 0.488 & {[0.481, 0.497]} & 0.498 & {[0.489, 0.505]} & 0.527 & {[0.520, 0.543]} & 528 \\ 
LLM-Ensemble & {\bfseries 0.576} & {[0.570, 0.583]} & {\bfseries 0.604} & {[0.597, 0.610]} & {\bfseries 0.593} & {[0.586, 0.607]} & {\bfseries 0.601} & {[0.595, 0.601]} & {\bfseries 0.634} & {[0.627, 0.638]} & {\bfseries 0.611} & {[0.597, 0.615]} & 555 \\ 
Omikuji & 0.499 & {[0.492, 0.521]} & 0.516 & {[0.512, 0.537]} & 0.531 & {[0.518, 0.560]} & 0.525 & {[0.513, 0.531]} & 0.561 & {[0.550, 0.566]} & 0.537 & {[0.525, 0.547]} & 534 \\ 
XR-Transformer & 0.530 & {[0.520, 0.545]} & 0.537 & {[0.529, 0.552]} & 0.581 & {[0.567, 0.595]} & 0.512 & {[0.502, 0.521]} & 0.531 & {[0.521, 0.540]} & 0.543 & {[0.533, 0.550]} & 586 \\ 
ZestXML & 0.459 & {[0.454, 0.470]} & 0.454 & {[0.449, 0.463]} & 0.522 & {[0.518, 0.538]} & 0.451 & {[0.440, 0.456]} & 0.448 & {[0.437, 0.457]} & 0.508 & {[0.493, 0.513]} & 542 \\ 
\bottomrule
\end{tabular*}
\caption{Graded Relevance Results in Tasks \texttt{Fulltext-30k} and \texttt{Book-Titles}.}
\label{tab:graded_rel}
\end{table*}

%% file: sections/06-08_final-sections.tex
\section{Limitations} \label{sec:limitations-and-future-work}

This benchmark focuses on scientific literature as indexed by the German
National Library, and therefore may not generalise to other domains or
document types (e.g., grey literature, newspapers, or popular science) or
other languages. The GND, while comprehensive, is tailored to specific
cultural and scholarly needs in German-speaking contexts and contains
idiosyncrasies that may affect transferability to other more specialised
authority files or ontologies. One particular point to note is, that this study makes
the simplifying assumption
to treat all entity types of the GND as the same. Arguably, named entities are
linguistically different from subject headings, when it comes to their assignment
as relevant labels to a text. The attribution to a text for named entities
is usually a binary decision, whereas subject headings may have varying degrees
of relevance for a particular text. This is not addressed within the XMLC-setting
of this study.   

Moreover, we were only able to compare a limited number of algorithms.
During the elapse of our project, new algorithms, such as Renee
\citep{Renee} or ViXML \citep{Ortego2026}, were released, offering potential advancement on the
state-of-the-art. The authors regret that these could not be investigated to full
extent. Not only the selection of algorithms but also the time spent
adapting the algorithms for our data, with hyper-parameter optimisation, is
subjective. One-Vs-All-Classifiers are considerably easier to parametrise
as opposed to transformer-based methods, which open up an almost
uncontrollable degree of potentially relevant hyperparameters%
\footnote{Consider, for example, the initial choice of pretrained encoder
  models.}. The authors have experimented with these hyperparameters to the
best of their knowledge, fully aware that poor performance may always occur, 
which can be attributed to a combination of the inability to control the 
investigated program and the inadequacy of the proposed algorithm.

Another class of hyperparameters influencing the study is related to the 
preprocessing of the texts. One such parameter is the selection, which parts of a
fulltext book should be processed, if not all. \citet{poley-etal-2025} have
experimented with Table-of-Contents (TOC) as a high quality source of training 
material. Future work should investigate this as an intermediate level of
information complexity between the \texttt{Book-Titles} and 
\texttt{Fulltext-30k} tasks. Given the length of the academic textbooks in our collection,
as illustrated in appendix~\ref{sec:length-distr}, dealing with full length text books has been 
computationally out of scope for this study. 

Other preprocessing parameters
concern the vectorisation and tokenisation for the TFIDF text representations. 
While these parameters may seem less dramatic, in contrast to text length, 
they do have direct impact on feature size and, thus, computational complexity
for the more traditional approaches. While initial ablation studies where 
conducted, the impact of the vectorisation was not studied in detail for all
methods.      

Further limitations of this study concern the graded relevance ratings. 
The resources of our subject experts for graded relevance ratings are
limited. We could not afford having all algorithms rated by the experts, so
we restricted this part to those methods most promising during preliminary
investigations that were suitable to be evaluated on both tasks. These time 
and resource constraints sadly excluded our final prototype KI-FSPrompt,
which would have been interesting to evaluate in the graded relevance setting, too.

Another note of warning relates to the LLM-based approaches: it is
impossible to capture \textit{the} state-of-the-art in LLM-based approaches, due to
the rapid advancements of methods and models in recent years. The
LLM-Ensemble and KI-FSPrompt used in this study are just snapshots in an
evolving landscape. Benchmarking LLM-based approaches would legitimise a new
benchmark study on its own.

An aspect that should have taken a prominent place in this study is the
evaluation of training and inference times. However, in the course of our
project, we have changed and switched hardware environments (between local
and HPC-based deployment) multiple times, which makes a consistent
comparison rather difficult.

Finally, there is a gap to bridge between the prototypes resulting from this
research and the requirements in the productive library process of automated
indexing. We recommend for library practitioners to construct their
productive systems around Annif, ultimately working toward increasing the
 availability of new backends in the Annif ecosystem.

\section{Conclusion and Future Work}

This study has contributed to the understanding of weaknesses and strengths
of the various algorithmic approaches to automated subject indexing. We see
that supervised and matching-based approaches, even in the age of
generative LLMs, still have their benefits and strengths. In terms of
absolute metrics, the supervised XMLC approach XR-Transformer shows best
results. In terms of vocabulary coverage and performance for less
frequent labels, the LLM-based approaches deliver higher quality, as confirmed 
by our graded expert ratings.

Future work should be directed towards refining the KI-FSPrompt pipeline,
which seems to be the most promising among our prototypes. In particular, more
research is needed on developing approaches that can also efficiently handle 
the content of longer documents, which cannot easily be processed as few-shot 
examples in a single prompt.

Finally, this study has focused on evaluating single backend algorithms. In
practice, productive systems should combine the most promising methods into
ensemble methods to achieve the best performance. Here, further research is
needed to identify component models for such ensembles, as well as methods for
ensembling that can successfully manage the challenges posed by large label spaces.

\section{Acknowledgements}

This work is a result of a research project at the German National Library
(DNB)\footnote{\url{https://www.dnb.de/ki-projekt}}. The project was funded
by the German Minister of State for Culture and the Media as part of the
national AI strategy. 
We extend our thanks to all of our subject librarians, whose countless hours 
of dedicated effort made the graded relevance ratings possible.
Furthermore, the authors gratefully acknowledge the computing time
made available to them on the high-performance computer at the NHR Center
of TU Dresden. This center is jointly supported by the Federal Ministry of
Research, Technology and Space of Germany and the state governments
participating in the NHR\footnote{\url{https://www.nhr-verein.de/unsere-partner}}.
Finally, we thank the anonymous reviewers for their kind and helpful feedback.

%% file: sections/App_datasets.tex
\appendix

\newpage

\section{Datasets}\label{sec:appendix}

\subsection{Sampling plan for graded relevance ratings}
\label{sec:appendix-sampling-plan}

The test set above consists of six data slices, each having the same distribution of
main DDC Subject Categories 
as the whole test set.
To ensure that the publications were not seen before, each algorithm was
tested on a distinct data slice.
Due to resource considerations, in the graded relevance scenario,
not all methods could be rated by subject experts. So only a subset of six methods could
be included. Also, only top 5 predictions per algorithm and task could be rated.

The test set was clear of gold
standard annotations before the beginning of the evaluation phase. Over a course
of three years, subject experts successively rated machine-based suggestions
for one algorithm per slice.\footnote{This
is to ensure that the document was only seen once by the subject expert.}
Afterwards they were asked to complete the metadata
record according to the rules of RSWK, also filling in missing aspects not discovered
by the algorithms.
The resulting test set has full gold standard annotations for all 4\,651 documents
and graded relevance ratings for one slice per method. Therefore, 
graded relevance metrics are based on distinct data slices per method, whereas
results for binary relevance metrics all rely on the full test set.

\subsection{Additional Statistics on Datasets}\label{sec:add-stats}

To avoid optimisation bias in the training and hyper-optimisation, we worked
according to a four-way-hold-out strategy: In addition to the training set
and test set described in section~\ref{sec:data}, we chose two validation
sets, \texttt{dev-train} and \texttt{dev-test}, for each task. 
The  \texttt{dev-train} splits were sampled from the same distribution as
the training data with
a stratified sampling technique by \citet{Merrillees2021}, separately for each task.
This ensures that the \texttt{dev-train} splits holds as many as possible of the labels also seen
during training and is therefore similar in composition to the training set.
For \texttt{dev-test}, however, the data is sampled in equal bins of the same 
scientific main subject categories also represented in the test set.
This ensures that all main subject categories
contribute with equal weight to the evaluation, even if the training data
is unevenly distributed over main subject categories.
The purpose of \texttt{dev-train} is to control learning and overfitting
during training. The purpose of \texttt{dev-test} is for hyper-parameter
optimisation, including threshold and limits to determine optimal precision-recall
balance as in section~\ref{sec:mediating-in-sample-bias}. 
The documents contained in the \texttt{dev-test}  split are the same for both tasks,
allowing direct comparison between the tasks. Table~\ref{tab:split-statistics}
shows further descriptive statistics on the development sets.

\begin{table}[h!]
  \centering
  \small
  \begin{tabular}{lccc}
    \hline
    \textbf{Task} & \textbf{Training} & \textbf{Dev-Train} & \textbf{Dev-Test} \\
    \hline
    \multicolumn{4}{l}{\textbf{Number of Documents}} \\
    \hline
    \texttt{Book-Titles} & 951\,104 & 73\,320 & 8\,415 \\
    \texttt{Fulltext-30k} & 167\,281 & 24\,315 & 8\,415 \\
    \hline
    \multicolumn{4}{l}{\textbf{Number of Labels}} \\
    \texttt{Book-Titles} & 193\,921 & 97\,894 & 11\,387 \\
    \texttt{Fulltext-30k} & 67\,757 & 40\,703 & 11\,387 \\
    \hline
    \hline
  \end{tabular}
  \caption{Dataset statistics for training and development sets.}
  \label{tab:split-statistics}
\end{table}

\subsection{Length Distribution of Training Data}
\label{sec:length-distr}

This appendix provides information on the length distribution of training texts
for the respective tasks. Table \ref{tab:length-distro} shows distributional 
statistics for \texttt{Book-Titles}, \texttt{Fulltext} and \texttt{Fulltext-30k}.
Here, \texttt{Fulltext} refers to the full length text books without cut off,
whereas \texttt{Fulltext-30k} applies a uniform cutoff. 

\input{tables/table_strlen.tex}

\begin{figure*}[!ht]
\centering
\begin{subfigure}[b]{0.95\textwidth}
    \includegraphics[width=\textwidth]{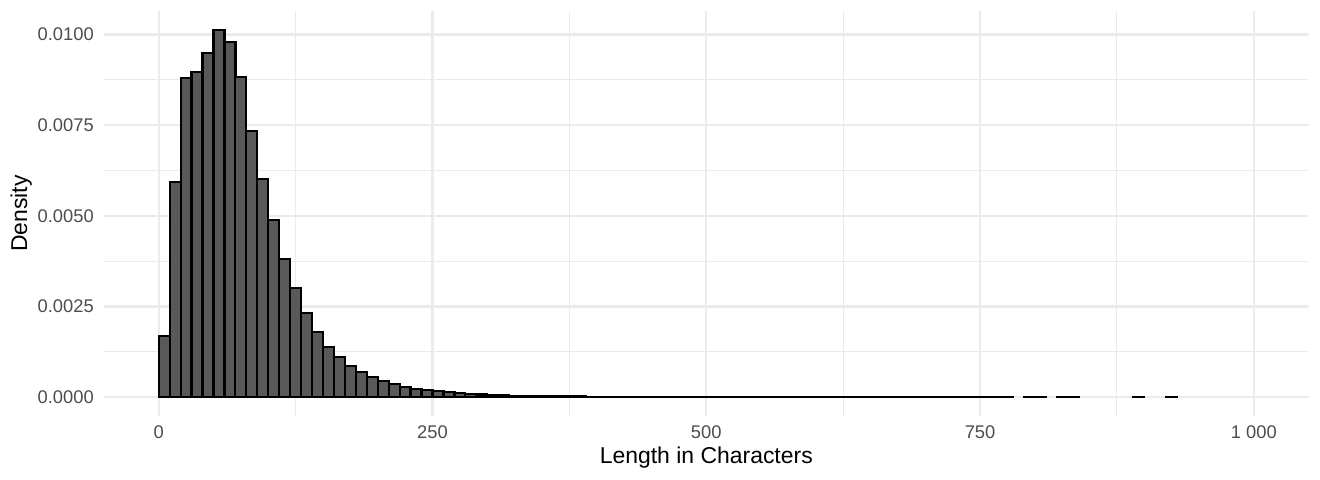}
    \caption{Task \texttt{Book-Titles}}
    \label{fig:length-distro-title}
\end{subfigure}
\begin{subfigure}[b]{0.95\textwidth}
    \includegraphics[width=\textwidth]{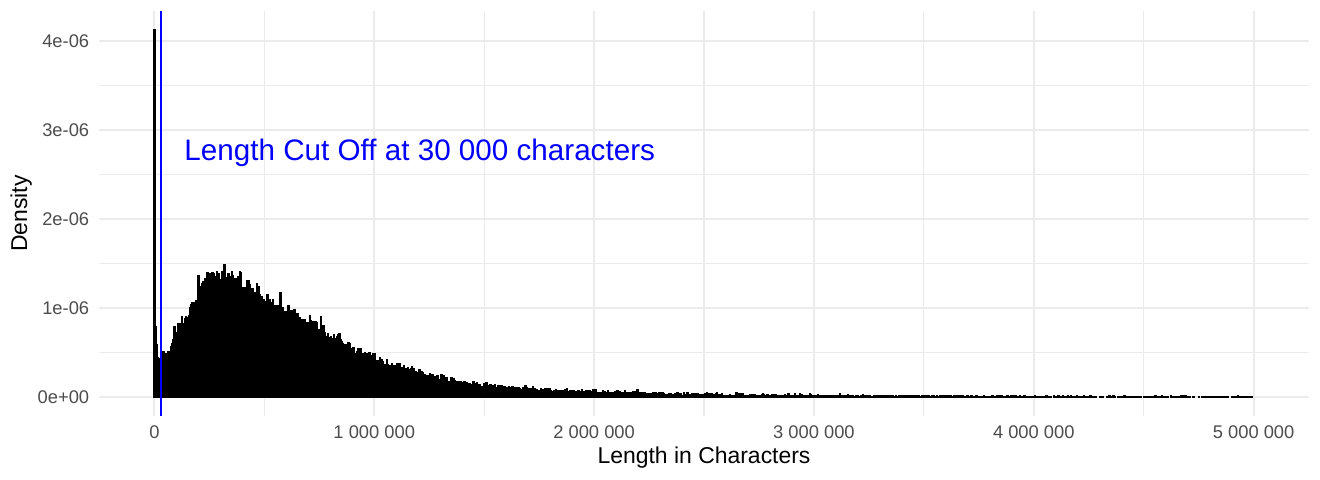}
    \caption{Task \texttt{Fulltext}}
    \label{fig:length-distro-ft}
\end{subfigure}
\caption{Distribution of training text length for tasks \texttt{Fulltext}
 and \texttt{Book-Titles}.}
\label{fig:length-distro}
\end{figure*}

%% file: tables/table_strlen.tex
\begin{table*}[t]
\small
\begin{tabular*}{\linewidth}{@{\extracolsep{\fill}}lrrrrrr}
\toprule
Task & Min & Q1 & Median & Mean & Q3 & Max \\ 
\midrule\addlinespace[2.5pt]
\texttt{Book-Titles} & 1 & 40 & 65 & 75 & 97 & 1 472 \\ 
\texttt{Fulltext} & 123 & 291 055 & 504 671 & 632 410 & 812 156 & 10 129 426 \\ 
\texttt{Fulltext-30k} & 123 & 29 993 & 29 997 & 29 565 & 29 999 & 30 000 \\ 
\bottomrule
\end{tabular*}
\caption{Distribution of text length (character count) for training data per task. \texttt{Fulltext} describes full texts without length cut off. }
\label{tab:length-distro}
\end{table*}

%% file: sections/App_examples.tex
\section{Examples}
\label{sec:app-examples}

Table~\ref{tab:example-books} shows two examples from our catalogue along with
their respective subject terms according to RSWK. In practice, subject 
librarians have access to the full document text, so it is not always
possible to infer all gold standard subject terms from a book title alone. 
Therefore, in the \texttt{Book-Titles} task, we must expect that there is a
certain amount of infeasible subject terms that can never be found by any 
of the methods. 

% cSpell:disable
\begin{table}[!h]
  \centering
  \scriptsize
  \begin{NiceTabular}{p{0.43\textwidth}}
  %\textbf{Example book title} & \textbf{Subject Terms}  \\
  \hline \\
  \makecell[l]{
    \textbf{Book title:} \\
    Wachstumsstrategien für Unternehmen: Wettbewerbsfähigkeit in \\
    disruptiven Zeiten sichern \\
    [\textit{Growth Strategies for Companies: Securing Competitiveness} \\
     \textit{in Disruptive Times}]
   } \\
  \makecell[r]{
    % \raggedleft
    \textbf{GND Subject Terms} \\
    Unternehmenswachstum \texttt{gnd-id:4078620-1} \\
    [\textit{Corporate Growth}] \\
    Strategisches Management \texttt{gnd-id:4124261-0} \\
     [\textit{Strategic Management}] \\
    Wettbewerbsfähigkeit \texttt{gnd-id:4065837-5} \\
     [\textit{Competitiveness}] \\
    Wachstumsstrategie \texttt{gnd-id:4520586-3} \\
    [\textit{Growth Strategy}]
  } \\
  \hline \\
  \makecell[l]{
    \textbf{Book title:} \\
    Forschungsmoral in der qualitativen Sozial- und Gesundheitsforschung\\
     Konflikte – Reflexion – Expertise\\
     [\textit{Scientific Integrity in Qualitative Social Sciences and Public}\\
      \textit{Health Research: Conflicts – Reflection – Expertise}]
  } \\
  \makecell[r]{
    % \raggedleft
    \textbf{GND Subject Terms} \\
    Empirische Sozialforschung  \texttt{gnd-id:4014606-6} \\
    [\textit{Empirical Social Science}]  \\
    Wissenschaftsethik \texttt{gnd-id:4066602-5} \\
     [\textit{Research Ethics}] \\
    Qualitative Methode   \texttt{gnd-id:4137346-7} \\
    [\textit{Qualitative Method}] \\
    Forschungsprozess \texttt{gnd-id:4155054-7} \\
     [\textit{Research Process}]  \\
    Gesundheitswissenschaften  \texttt{gnd-id:4249464-3} \\
    [\textit{Public Health Research}] \\
    Moralisches Handeln  \texttt{gnd-id:4535360-8} \\
    [\textit{Moral Action}] \\
  } \\
  \hline
  \end{NiceTabular}
  \caption{Subject Terms and Labels for Example Books.}
  \label{tab:example-books}
\end{table}

%% file: sections/App_results.tex
\section{Additional Results}
\label{sec:app-results}

In this appendix section we provide additional data analysis to give a better
understanding of the results presented in the main results section~\ref{sec:results}.

\subsection{Precision-Recall Analysis}

\begin{figure}[!ht]
  \centering
  \includegraphics[width=0.45\textwidth]{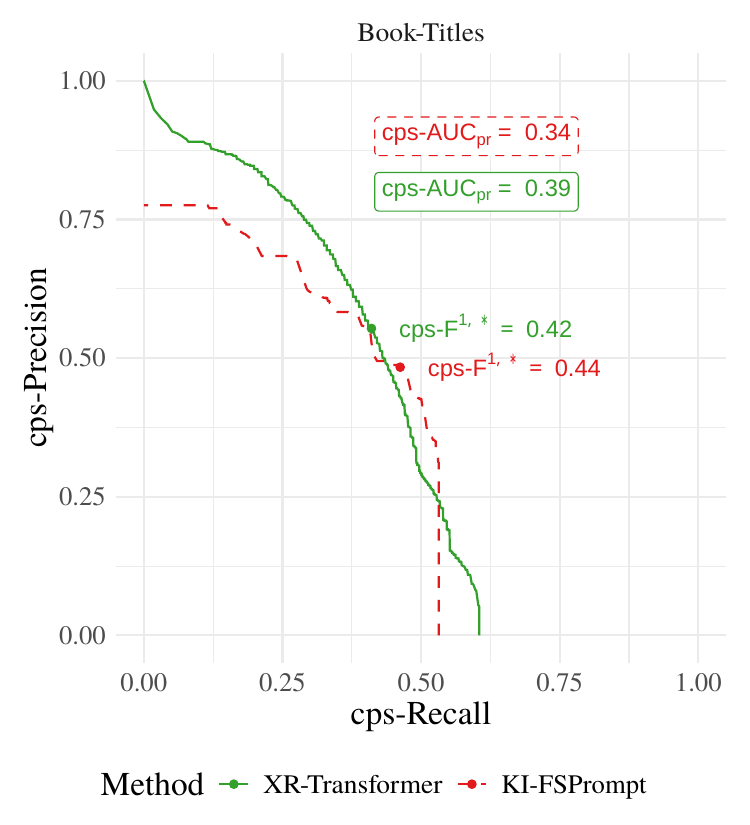}
  \caption{Conditionally propensity scored precision-recall curves for methods
  XR-Transformer and KI-FSPrompt, evaluated on the \texttt{Book-Titles} task.
  Curves are computed by varying the decision threshold and reporting resulting document-averaged
  precision and recall.}
  \label{fig:pr_curve_1}
\end{figure}

 Figure~\ref{fig:pr_curve_1} highlights that, while KI-FSPrompt achieves
  the highest point estimate for $\textup{cps}-F^{1,*}$, XR-Transformer maintains a
  higher $\textup{cps-AUC}_{\text{pr}}$. KI-FSPrompt only produces short lists
  of subject suggestions per record, varying in length between 3-10 subject
  terms. XR-Transformer obtains confidence scores for all labels, enabling
  arbitrary length of subject suggestion lists. 

\subsection{Performance by label frequency}
\label{sec:app-label-freq}

\begin{figure*}[!ht]
\centering
\begin{subfigure}[b]{0.95\textwidth}
    \includegraphics[width=\textwidth]{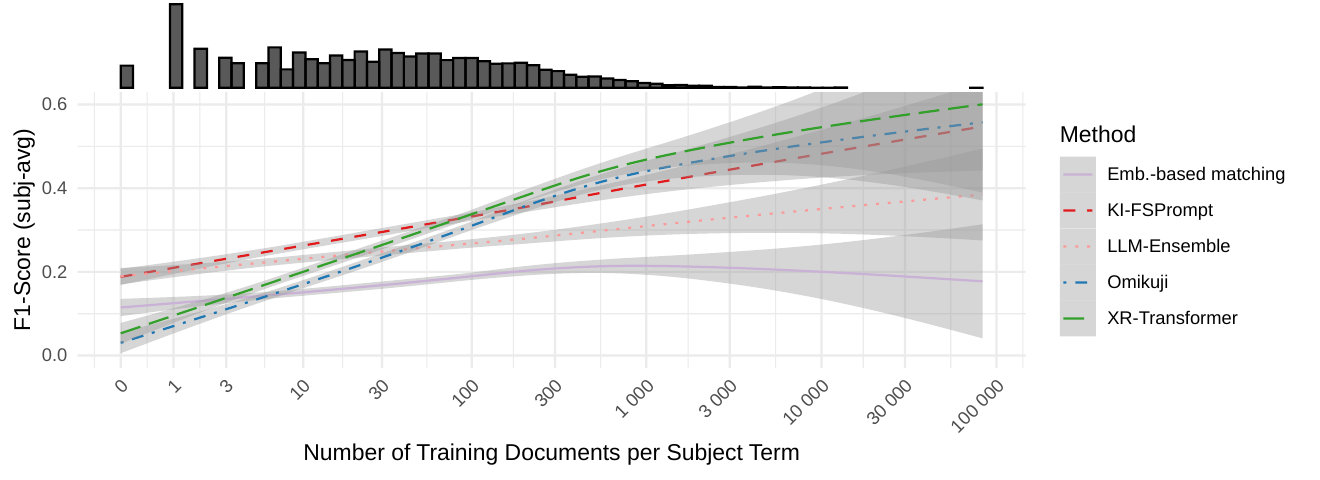}
    \caption{Task \texttt{Book-Titles}}
    \label{fig:res-by-freq-title}
\end{subfigure}
\begin{subfigure}[b]{0.95\textwidth}
    \includegraphics[width=\textwidth]{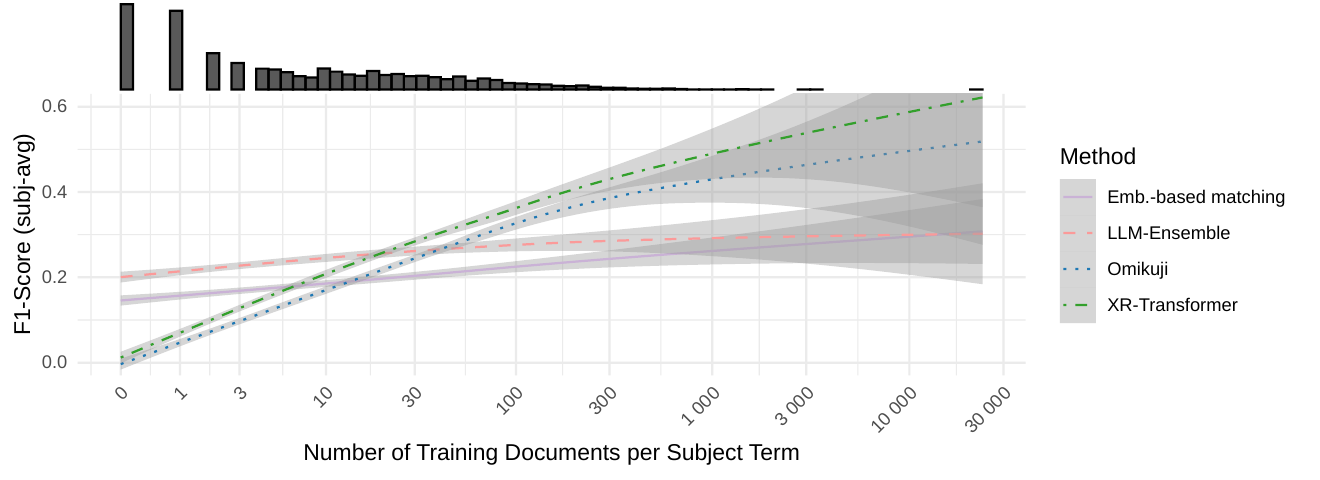}
    \caption{Task \texttt{Fulltext-30k}}
    \label{fig:res-by-freq-ft}
\end{subfigure}
\caption{Subject-average $F^1$-Score plotted as a function of label frequency.}
\label{fig:res-by-freq}
\end{figure*}

The following additional analysis compares the subject average performance
for selected methods as a function of label frequency in the training dataset.
For any given label in \texttt{GND-204K} we compute the $F^1$-score in analogy
to eq.~(\ref{eq:f1-def}). From these label-wise scores we estimate a continuous
spline with the  R package \texttt{mgcv} \citep{MGCV} as a function of training frequency.
To illustrate how many data points support any region of the spline, we include
the marginal histogram that plots the distribution of labels in the test set.
Figure~\ref{fig:res-by-freq-title} shows the results for task \texttt{Book-Titles}
and figure~\ref{fig:res-by-freq-ft} for task \texttt{Fulltext-30k}.
Note that the marginal distribution differs between the two tasks, even though
both tasks share the same test set. The training dataset for task \texttt{Book-Titles}
is much larger than for task \texttt{Fulltext-30k}. Hence, labels that appear
as few- or zero-shot for \texttt{Fulltext-30k} move further to the range
of common labels in \texttt{Book-Titles}.

A key takeaway from this analysis is that the LLM-based methods provide a
wider coverage of the vocabulary, whereas supervised XMLC methods excel
mainly in the domain of frequent labels. LLM-Ensemble and EBM are not trained,
 so the quality does not increase notably for
more frequent labels. However, KI-FSPrompt, leveraging the training dataset
to collect similar documents as few-shot examples, shows a similar increase
in quality with training frequency as the supervised models.

\subsection{Graded Relevance Ratings}

\begin{figure}[htb]
    \centering
    \includegraphics[width=0.45\textwidth]{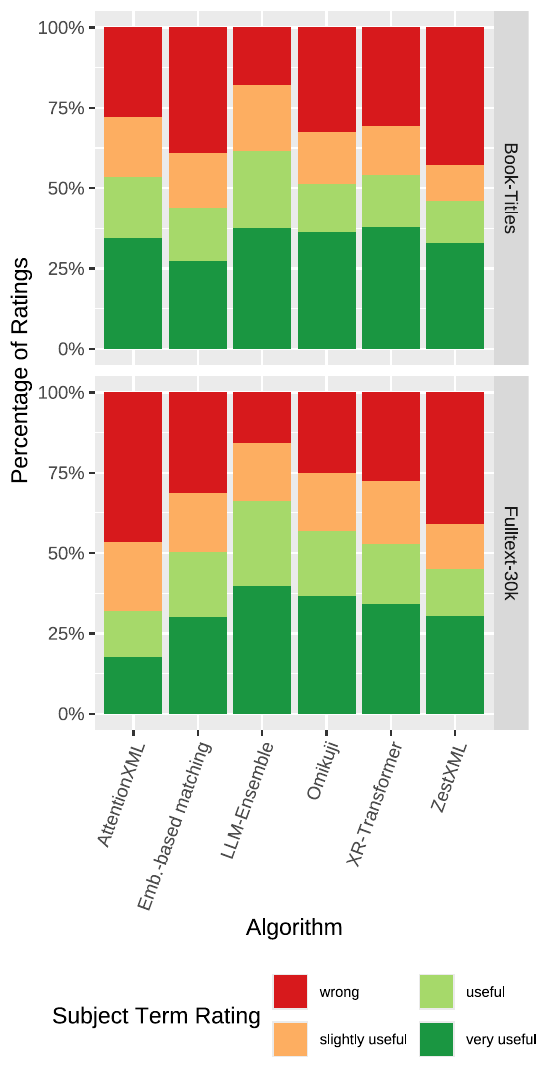}
    \caption{Relative distribution of graded relevance ratings per method
      for top 5 predictions per record.}
    \label{fig:qm-ratings}
\end{figure}

Figure~\ref{fig:qm-ratings} shows the relative distribution of graded relevance
ratings as conducted by our subject experts.

%% file: sections/App_metrics.tex
\section{Full metric definitions}
\label{sec:app-metrics}

This benchmark study uses some non-standard metric definitions. To make
the precise calculations transparent, we include the full definitions of
all metrics in this appendix.

\subsection{Binary Relevance Scenario}

Let $\textbf{y}_d = \{y_{d,l}\}_{l=1}^L \in \{0,1\}^L$ be the binary
indicator for gold standard subject terms of
document $d = 1, \ldots, N_{\textup{test}}$ with labels $l = 1, \ldots, L$,
according to the subject experts. Here,
$L$ denotes the size of our vocabulary (i.e. the label space) and
$N_{\textup{test}}$ is the number of documents in our test set. Likewise,
let $\hat{\textbf{y}}_d = \{\hat{y}_{d,l}\}_{l=1}^L \in \{0,1\}^L$ denote an
algorithm's subject predictions with confidence levels
$\hat{\textbf{c}}_d = \{ \hat{c}_{d,l}\}_{l=1}^L \in \left[0,1\right]^L$.
Let
\[
  \textup{Thres}_t (\hat{\textbf{y}}_d) := \{l ~|~ \hat{c}_{d,l} >= t\}
\]
be the set of label predictions above a threshold $t$ and
\[
  \textup{Top}_k (\hat{\textbf{y}}_d) := \{l~|~\#\{ l'~|~\hat{c}_{d,l'} > \hat{c}_{d,l} \} < k    \}
\]
be the top k ranked labels for some integer $k$.
We define $\textup{tp}_k$ (true positives), $\textup{fp}_k$ (false positives) and
$\textup{fn}_k$ (false negatives) as follows:
\begin{align}
  \label{eq:bin-rel-tp}
\tpk\left(\textbf{y}_d, \hat{\textbf{y}}_d\right) &:=
  \sum_{l \in \textup{Top}_k (\hat{\textbf{y}}_d)} y_{l,d} \\
  \label{eq:bin-rel-fp}
\textup{fp}_k \left(\textbf{y}_d, \hat{\textbf{y}}_d\right) &:=
  \sum_{l \in \textup{Top}_k (\hat{\textbf{y}}_d)} \left(1 - y_{l,d}\right) \\
  \label{eq:bin-rel-fn}
\textup{fn}_k\left(\textbf{y}_d, \hat{\textbf{y}}_d\right) &:=
   \sum_{l \notin \textup{Top}_k (\hat{\textbf{y}}_d)} y_{l,d}
\end{align}
and use the standard definitions for document-wise precision and recall
\begin{equation}
  \label{eq:prec_k}
  \textup{Prec}_k\left(\textbf{y}_d, \hat{\textbf{y}}_d\right) := \frac{\textup{tp}_k}{\textup{tp}_k + \textup{fp}_k}
\end{equation}
and
\begin{equation}
  \label{eq:rec_k}
  \textup{Rec}_k\left(\textbf{y}_d, \hat{\textbf{y}}_d\right) := \frac{\textup{tp}_k}{\textup{tp}_k + \textup{fn}_k}.
\end{equation}
Likewise, we have  $\textup{Prec}_{k,t}$ and  $\textup{Rec}_{k,t}$,
where the sum is built over
$l \in \textup{Top}_k (\hat{\textbf{y}}_d) \cap \textup{Thres}_t (\hat{\textbf{y}}_d)$,
applying threshold and limit simultaneously.
The document-wise $F^1$-Score is defined as\footnote{This is equal to the
harmonic mean of precision and recall whenever both are non-zero.}
\begin{equation}
  \label{eq:f1-def}
  F^{1}_{t,k}  (\textbf{y}_d, \hat{\textbf{y}}_d) :=
    \frac{2 \cdot  \textup{tp}_k}{
      2 \cdot \textup{tp}_k + \textup{fp}_k + \textup{fn}_k
    }
\end{equation}
The document-average (doc-avg) Precision is defined as
\begin{equation}
  \textup{Prec}_{k,t} := \frac{1}{N_{\textup{docs}}} \sum_{d=1}^{N_{\textup{docs}}} \textup{Prec}_{k,t} (\textbf{y}_d, \hat{\textbf{y}}_d)
\end{equation}
and analogously for Recall and F-score.
The precision-recall curve $\mathcal{C}_{\textup{pr}}$ is defined pointwise for a given level of recall $R$
\begin{equation}
  \mathcal{C}_{\textup{pr}}(R) := \max_{k,t} \left\{ \textup{Prec}_{k,t}~|~\textup{Rec}_{k,t} >= R \right\}
\end{equation}
and the area under the precision-recall curve is defined as
\begin{equation}
  \textup{AUC}_\textup{pr} := \int_0^1 \mathcal{C}_{\textup{pr}}(r) \textup{d}r
\end{equation}
Note that the pr curve here is built on document-average precision and recall,
and the search space is over limits $k$ and thresholds $t$ at any given point
of the pr curve.
All metrics are calculated using our public R package \texttt{CASIMiR}.\footnote{\url{https://github.com/deutsche-nationalbibliothek/casimir\allowbreak}}

\subsection{Mediating In-Sample Bias for Estimation of Optimal F-Score}
\label{sec:mediating-in-sample-bias}

To avoid in-sample bias for $F^{1,*}$, we calculate the optimal thresholds
and limits with pr curves on a separate \texttt{dev-test}-set which follows
a similar distribution as \texttt{test}, i.e. has a similar composition across
subject categories. See also section~\ref{sec:add-stats}.

\subsection{Metrics with conditional label weights}

In the above metrics, false positives ($\textup{fp}$), true positives ($\textup{tp}$)
 and false negatives ($\textup{fn}$)
enter the formula for precision and recall using the standard formulae (\ref{eq:prec_k})
and (\ref{eq:rec_k}).
To study the performance in the long tail of our vocabulary, we introduce
a specific cost scheme:

\begin{table}[!h]
  \label{tab:cost-scheme}
  \small
  \begin{tabular}{|l|c|c|}
    \hline
     & \textbf{Gold standard y} & \textbf{Gold standard n} \\
    \hline
    \textbf{prediction y} & $C_{\textup{tp}}\cdot \textup{tp}$ & $C_{\textup{fp}}\cdot\textup{fp}$ \\
    \hline
    \textbf{prediction n} & $C_{\textup{fn}}\cdot\textup{fn}$ & omitted \\
    \hline
  \end{tabular}
\end{table}

\noindent Here $C_{\textup{tp}}$, $C_{\textup{fp}}$ and $C_{\textup{fn}}$ are the costs
for true positives, false positives and false negatives, respectively. Such
cost factors can be used to introduce preferences in the evaluation metrics.
Our special focus is on rare subjects: We argue that rare subjects are more
informative and, thus, more costly to be missed out and hence the cost of a
false negative should be inversely proportional to
the number of training documents for that subject. The same holds true for
true positives, i.e. it should be encouraged getting rare subjects right.
However, getting rare subjects wrong is just as bad as getting a frequent subject
wrong. Therefore, the cost of false positives is assumed to be the same for all
subjects, independently of their frequency of occurrence.
As weights we propose the inverse propensity scores according to \citet{PfastreXML}:
$$C_{\textup{tp}} = C_{\textup{fn}} := w_{l} = 1 + C\cdot (n_{l} + B)^{-A},$$
where $C =  (\log N_{\textup{train}} - 1)\cdot (B + 1)^A$, $N_{\textup{train}}$
is the total number of training documents, $n_{l}$ the frequency of occurrence of $l$
within the train set and $A, B$ are free parameters.
\citet{PfastreXML} propose $A = 0.55$ and $B = 1.5$.
For the false positive cost, $C_{\textup{fp}}$, we chose a label independent
cost calculated as the mean of all $w_{l}$ across the test set:
$$
  C_{\textup{fp}} := \frac{1}{H}\sum_{d=1}^{N_\textup{test}} \sum_{l=1}^L w_l y_{l,d}
$$
with the normalisation constant:
$$
H:= \sum_{d=1}^{N_\textup{test}} \sum_{l=1}^L y_{l,d},
$$
which is the sum of document-label pairs in the test set according to the gold standard.
With the definitions for gold standard labels $\textbf{y}_d$ and machine-based predictions
$\hat{\textbf{y}}_d$ as before, we define
$$
\tilde{\textup{tp}}_k\left(\textbf{y}_d, \hat{\textbf{y}}_d\right) :=
  \sum_{l \in \textup{Top}_k (\hat{\textbf{y}}_d)} w_l \cdot y_{l,d}
$$

$$
\tilde{\textup{fp}}_k \left(\textbf{y}_d, \hat{\textbf{y}}_d\right) :=
  \sum_{l \in \textup{Top}_k (\hat{\textbf{y}}_d)} C_{\textup{fp}} \cdot \left(1 - y_{l,d}\right)
$$

$$
\tilde{\textup{fn}}_k\left(\textbf{y}_d, \hat{\textbf{y}}_d\right) :=
   \sum_{l \notin \textup{Top}_k (\hat{\textbf{y}}_d)} w_l \cdot y_{l,d}
$$

Finally, we define the \textbf{conditionally propensity scored precision} for a
document $d$ as:

\begin{equation}
  \textup{cps-Prec}_k  :=
    \frac{
      \tilde{\textup{tp}}_k
    }{
      \tilde{\textup{fp}}_k  +
        \tilde{\textup{tp}}_k
    }
\end{equation}
and  \textbf{conditionally propensity scored recall}:
\begin{equation}
  \textup{cps-Rec}_k  :=
    \frac{
      \tilde{\textup{tp}}_k
    }{
      \tilde{\textup{fn}}_k  +
        \tilde{\textup{tp}}_k
    }
\end{equation}
and analogously $\textup{cps-}F^1_k$ as in eq.~\ref{eq:f1-def}.
Document-average versions of $\textup{cps-Rec}_k$ and $\textup{cps-Prec}_k$ as
well as the conditionally propensity scored $F^1$-Score are defined as above.
Also, $\textup{cps-Rec}_k$ and $\textup{cps-Prec}_k$ can be generalised to the
case where not only a uniform limit $k$ is applied, but also a uniform
threshold $t$ is used to cut off machine-based predictions. Finally, varying
limits and thresholds give rise to a conditionally propensity scored
precision-recall curve
$\textup{cps-}\mathcal{C}_\textup{pr}$ with its respective conditionally propensity scored
area under the curve $\textup{cps-AUC}_\textup{pr}$.
This metrics emphasises an algorithm performance with respect
to its long-tail performance across the entire ranking.

\subsection{Generalised Precision and Recall}
\label{sec:app-graded-relevance}

For the definitions of generalised precision $\textup{g-Prec}$ and generalised
recall $\textup{g-Rec}$ we follow \citet{Kekalainen2002}. Let
$r_{l,d}\in\left[0,1\right]$ be the graded
relevance rating of label $l$ for document $d$ and
$\textbf{r}_d = \{r_{l,d}\} \in\left[0,1\right]^L$.\footnote{Relevance ratings $r_{l,d}>1$ should be normalised to the standard interval.}
Let us define
$$
\Delta_\textup{rel}\left(\textbf{y}_d, \hat{\textbf{y}}_d, \textbf{r}_d \right):= \sum_{l \in \textup{Top}_k (\hat{\textbf{y}}_d)} r_{l,d}\cdot \left(1 - y_{l,d}\right),
$$
as the sum of relevance scores for all false positive Top-k suggestions.
With this definition, $\textup{g-Prec}$ and $\textup{g-Rec}$ can be expressed using
the same definitions for $\textup{tp}_k, \textup{fp}_k$ and $\textup{fn}_k$
as in eq. (\ref{eq:bin-rel-tp})-(\ref{eq:bin-rel-fn}). With the correction term
$\Delta_\textup{rel}$, we can write the document-wise generalised precision as:\footnote{
Here, we have omitted that $tp$, $fp$, $fn$ and $\Delta_\textup{rel}$ also take
the document dependent arguments  $\textbf{r}_d,\textbf{y}_d$ and $\hat{\textbf{y}}_d$
}
\begin{equation}
  \label{eq:g-prec_k}
  \textup{g-Prec}_k\left(\textbf{y}_d, \hat{\textbf{y}}_d, \textbf{r}_d \right) :=
    \frac{\textup{tp}_k + \Delta_\textup{rel}}{\textup{tp}_k + \textup{fp}_k}.
\end{equation}
Likewise, the document-wise generalised recall is:
\begin{equation}
  \label{eq:g-rec_k}
  \textup{g-Rec}_k\left(\textbf{y}_d, \hat{\textbf{y}}_d, \textbf{r}_d \right) :=
     \frac{\textup{tp}_k + \Delta_\textup{rel}}{\textup{tp}_k + \textup{fp}_k + \Delta_\textup{rel}}.
\end{equation}
From these definitions we immediately derive document-average
$\textup{g-Prec}_k$ and $\textup{g-Rec}_k$, which are the metrics reported
in section \ref{sec:results-grad-rel}.
Writing $\textup{g-Prec}_k$ and $\textup{g-Rec}_k$ in this form, we observe
that $\textup{g-Prec}_k > \textup{Prec}_k$ and $\textup{g-Rec}_k > \textup{Rec}_k$
whenever $\Delta_\textup{rel} > 0$.